\documentstyle[prl,aps,floats,psfig]{revtex}

\begin{document}
\twocolumn[
\hsize\textwidth\columnwidth\hsize\csname@twocolumnfalse\endcsname

\draft
\title{
Phonon `notches' in $a$-$b$-plane optical conductivity \\
of high-$T_c$ superconductors
}
\author{Vladimir N. Kostur\cite{na}}
\address{
Center for Superconductivity Research, Department of Physics\\
University of Maryland, College Park, MD 20742
}
\date{{\bf E-print cond-mat/9507142}; received July 31, 1995}
\maketitle
%
%
%

\begin{abstract}
It is shown that a correlation between the positions of
the $c$-axis longitudinal
optic ($LO_c$) phonons and ``notch''-like structures in the
$a$-$b$
plane  conductivity  of high-$T_c$ superconductors
results from phonon-mediated interaction
between  electrons in different
layers.
It is found that the relative size of the notches
depends on
 $\lambda_{ph}(\Omega_{ph}/\gamma_{ph})$, where
$\lambda_{ph}$, $\Omega_{ph}$ and $\gamma_{ph}$ are the
effective coupling strength,
the frequency and the width of the optical phonon which is responsible
for the notch. Even for $\lambda_{ph}\approx 0.01$ the effect can be large
if the phonon is very sharp.
\end{abstract}

\pacs{PACS numbers: 78.30.Er, 63.20.Kr, 72.10.Di, 74.70.Vy}
]

The striking correlation between the positions of the $c$-axis longitudinal
optic ($LO_c$) phonons and ``notch''-like absorption structures
in the $a$-$b$-
plane conductivity $\sigma_{ab}({\bf q}^{0},\omega)$
was found by Reedyk
and Timusk \cite{Reedyk1}
for most of the high-$T_c$ copper-oxide superconductors.
It was shown \cite{Reedyk1}  that such antiresonant
features are not present in infrared measurements if
the wave vector of light ${\bf q}^{0}$ is perpendicular to the $c$-axis.
These results strongly suggested that the origin of ``notch''-like
structures in $\sigma_{ab}$ is in the interaction of the $a$-$b$-plane
electronic continuum with $LO_c$ phonons \cite{Timusk1,Timusk2},
rather than in
superconducting gap(s) \cite{Tanner}. This was further corroborated by
Reedyk {\em et al.\ } \cite{Reedyk2} in a detailed analysis of their
$a$-$b$-plane reflectivity measurements
on the $Pb_{2}Sr_{2}LCu_{3}O_{8}$-series ($L=Y,Dy,Eu,Sm,Nd$, and $Pr$),
where an intrinsically low carrier concentration
enables one to follow closely
the progression of the phonon features as one goes from an insulating to
a metallic regime. While the $a$-$b$-plane conductivity of the
insulating compounds below $700~cm^{-1}$
is dominated by a series of infrared-active phonon peaks,
the phonon peaks have by and large
disappeared in the conductivity of metallic compounds due to screening and
several antiresonant ``notch''-like features appear at the positions of
$LO_c$ phonons.

Nevertheless, a microscopic understanding of these antiresonant features is
lacking. The models presented in \cite{Timusk1,Timusk2} are purely
phenomenological and
the connection that was claimed \cite{Reedyk1,Timusk2,Reedyk2} to exist
to the microscopic theory by M. J. Rice  \cite{Rice}
of the optical properties of the {\em quasi-one-dimensional}
organic conductors is--
at best--unclear.

Here a microscopic mechanism which accounts for the
presence of antiresonant
features in $a$-$b$-plane optical conductivity when the wave vector of
light ${\bf q}^{0}$ is parallel to the $c$-axis, and for their absence when
${\bf q}^{0}$ is perpendicular to the $c$-axis is proposed.
(While in infrared ${\bf q}^{0}$
is small, it is finite-of the order $1/\delta$, where $\delta$ is the
penetration depth of light in the material.)
The contribution to the conductivity arising from the Fano process \cite{Fano}
depicted in Fig.\ 1(a) is considered.
\begin{figure}
\centerline{\psfig{file=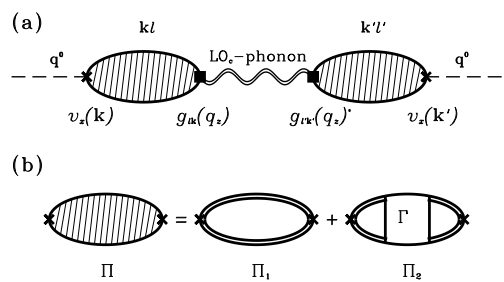,width=0.95\linewidth}}
\caption{(a) The ``one-phonon-assisted''
 contribution to the current-current correlation function.
The dashed line represents a
photon and the wavy line represents a dressed
$LO_c$-phonon (${\bf q} \| c$).
$g_{l{\bf k}}( q_z)$
is the electron-phonon coupling matrix element.
(b) Diagrammatic representation of the
 current-current correlation function. $\Gamma$ is a four-point vertex
due to interactions in the $a$-$b$-plane.}
\end{figure}
An infrared photon polarized along the $a$-$b$-plane
(a dashed line in Fig.\ 1(a)) creates a
particle-hole pair. In what follows the electric field vector $\bf E$
is taken to be along $x$-direction (parallel to $a$-axis), Fig.~1(a).
The particle and the hole experience all possible
uncorrelated (the term $\Pi_{1}$ in Fig.\ 1(b)) and correlated
(the term $\Pi_{2}$ in Fig.\ 1(b)) scattering events.
 The particle and the hole
could recombine and produce a phonon (the wavy line in Fig.\ 1(a)).
We assume that  interaction between electrons in the layer labeled $l$
and $LO_c$-phonon is described by Fr\" ohlich-like  term\cite{KosturMitrovic}
\begin{equation}
H_{e-LO_c}=\sum_{l{\bf k}\sigma} c_{l{\bf k}\sigma}^{\dagger}
c_{l{\bf k}\sigma} \sum_{q_z} g_{l{\bf k}}(q_{z})\ \big[
b_{q_z} + b_{-q_z}^{\dagger}\big]\>,
\end{equation}
where
$l$ labels the layers stacked along $z$-direction (parallel to the
$c$-axis),
$\bf k$ is the momentum in the $a$-$b$-plane and  $\sigma$ is the
spin index.
Here, it is taken into account that
$LO_c$-phonon can carry the momentum only in $c$-direction.
The bare electron-phonon matrix element can be written in the following form
\cite{AllenMitrovic}
\begin{eqnarray}
g_{l{\bf k}}^{(0)}(q_{z}) & = & - \sum_{\nu}\frac{1}{\sqrt{2NM_{\nu}
\omega_{q_{z}}^{LO_c}}}\ e^{i q_z Z_{\nu}^{0}}
\nonumber\\
& \times & \sum_{\bf K}
\langle l,{\bf k}|{\bf e}_{LO_c}(q_z)  i
\nabla U({\bf r - R_{\nu}^{0}})|l,{\bf k + K}\rangle
\>.
\end{eqnarray}
Here $M_{\nu}$, $\bf R_{\nu}^{0}$ and
$U({\bf r - R_{\nu}^{0}})$ are, respectively,
the mass, position and potential of $\nu$-th atom.
$N$ is the number of atoms.
$\omega_{q_z}^{LO_c} \approx \Omega_{ph}$ is $LO_c$-phonon frequency and
$\bf K$ is a reciprocal lattice vector.
The polarization vector ${\bf e}_{LO_c}(q_z)$ is
parallel to $c$-axis.
The contribution from electron-$LO_c$-phonon
interaction to current-current correlation function
is proportional to
\begin{equation}
\delta \Pi(q_{z}^{0}, \omega) = \sum_{ll'} \delta \Pi_{ll'}(\omega)\ e^{i
q_{z}^{0}
(z_{l} -z_{l'})}\>,
\end{equation}
where the contribution from the diagram
with one-phonon line which connects
two particle-hole polarization bubbles is (see Fig.~1(a))
\begin{eqnarray}
\delta \Pi_{ll'}(\omega)& = & \frac{\Pi_{l}(\omega)
\Pi_{l'}(\omega)}{N(0)}
 \sum_{q_{z}}
D(\Omega_{ph}, \omega)
\sum_{{\bf k}\in FS} e v_{x}({\bf k})
\nonumber \\ & \times & g_{l{\bf k}}(q_z)
\sum_{{\bf k'}\in FS} e v_{x}({\bf k'})\ g_{l'{\bf k'}}(q_z)^{*}
\>.
\end{eqnarray}
Here $N(0)$ is the density of states at the Fermi level,
 $v_{x}({\bf k})$ is the electron velocity on the Fermi surface
coming from the current vertex,
$\Pi_{l}(\omega)$
is a polarization operator for $l$th layer and
$D$ is the $LO_c$-phonon propagator.  The sums over $\bf k,k'$
are restricted to the Fermi surface (FS).
For simplicity, it is  assumed that
$\Pi_{l}$ is independent
on $l$.
At this stage I would like to emphasize the role of layered
structure.  In three-dimensional case the average
of velocity times electron-phonon matrix element
over the Fermi surface is equal to
zero.
In the case under consideration
I propose an assumption  that the quantity
\begin{equation}
\lambda_{ll'}({\bf k,k'}) = N(0) \sum_{q_z}\ g_{l{\bf k}}(q_z)
g_{l'{\bf k'}}(q_z)^{*}
\end{equation}
is sharply peaked at ${\bf k} \approx {\bf k'}$.
This  means that the electrons from different
layers interact appreciably via   exchange of the
$LO_c$-phonon when their  momenta are nearly equal.
With this assumption
one finds a {\em non-zero} contribution at $q_{z}^{0}\neq 0$
\begin{eqnarray}
\delta\sigma(\omega) & = & \sigma({\bf q}^{0}\| c, \omega)-
\sigma({\bf q}^{0}\bot c, \omega)
\\
& = & - \frac{\lambda_{ph}(q_{z}^{0})\omega_{p}^{2}}{4\pi\omega}
\ \Pi(\omega)^2  D(\Omega_{ph},\omega)\>.
\end{eqnarray}
Here $\omega_{p}$ is the plasma frequency and
\begin{eqnarray}
\lambda_{ph}(q_{z}^{0}) & = & \sum_{{\bf k,k'}\in FS}
v_{x}({\bf k}) v_{x}({\bf k'})
\sum_{ll'} e^{iq_{z}^{0}(z_{l} - z_{l'})}
\nonumber \\ & \times  &
\lambda_{ll'}({\bf k},{\bf k}')/\sum_{{\bf k}\in FS}
v_{x}({\bf k})^{2}\>.
\end{eqnarray}
In the case when ${\bf q}^{0}$ is perpendicular to the $c$-axis
$\lambda_{ph}(q_{z}^{0}=0)=0$, because the electron-phonon matrix
element of the $LO_c$-phonon is proportional to
${\bf e}_{LO_c}\cdot \nabla \propto \partial/\partial z_{l}$, and
it vanishes after summation over $l$.

The experimentally measured values of the electron-phonon coupling
parameter for a particular $c$-axis phonon branch, e.~g.
$B_{1u}(z)$ optical modes,  are
estimated to be $0.01$-$0.03$ \cite{Thomsen}, in agreement with calculated
values \cite{Rodriguez}. The effective value of $\lambda_{ph}(q_{z}^{0})$ in
Eq.~(6) can be smaller (not much if $\lambda_{ll'}({\bf k,k'})$
is sharply peaked at $\bf k=k'$).
However, as t is  shown below, the maximum relative size of the
contribution from this process is about $\lambda_{ph}(q_{z}^{0})
\Omega_{ph}/\gamma_{ph}$,
where $\Omega_{ph}$ and $\gamma_{ph}$ are the frequency and the width of
the phonon, respectively.
The experimentally measured values of $\Omega_{ph}/\gamma_{ph}$ for $LO_c$
phonons are in the range $10$-$200$ \cite{Thomsen}. The band
structure calculations give nearly the same or larger values for this ratio
\cite{Rodriguez}.
Thus, if the $a$-$b$ plane $LO$ phonons are  sharp their contribution
through the Fano process can  be considerable
even for {\em small} values of $\lambda_{ph}(q_{z}^{0})$
as is found experimentally \cite{Reedyk1}.

The contribution to the real part of the $a$-$b$-plane conductivity
is given by
\begin{eqnarray}
\delta\sigma_1(\omega)  =  -\frac{\omega_{p}^{2}}{4\pi\omega}
\Bigg[\left[ \Big( Re\Pi(\omega) \Big)^2-\Big( Im\Pi(\omega) \Big)^2\right]
\phantom{\>.}
\nonumber\\
 \times
Im D(\Omega_{ph},\omega)
+2Re\Pi(\omega)Im\Pi(\omega)
Re D(\Omega_{ph},\omega)\Bigg]\>.
\end{eqnarray}
The propagator $D$ for optical phonon can be modeled  by a Lorentzian
\begin{equation}
D(\Omega_{ph},\omega)=\frac{\Omega_{ph}^2}
{\omega^2-\Omega_{ph}^2+i\gamma_{ph}\omega}\>.
\end{equation}
The function $\Pi(\omega)$
is connected to the scattering processes of the charge carriers.
To  estimate the order
of magnitude of $\delta\sigma_1(\omega\approx \Omega_{ph})$
I use the following approximate
form for $\Pi$ in the normal state \cite{Allen,Gotze}
\begin{equation}
\Pi(\omega)=\frac{\omega}{(1+\bar{\lambda}(\omega))\ \omega
   + i\tau^{-1}(\omega)}\>.
\end{equation}
Here $\tau^{-1}(\omega)$ is the scattering rate of the charge carriers and
$(1+\bar{\lambda}(\omega))$ is an effective mass renormalization parameter.
By using the Eq.~(9) one can obtain an estimate at $\omega=\Omega_{ph}$
\begin{equation}
\delta\sigma_1(\Omega_{ph})=-\frac{\omega_{p}^{2}}{4\pi}
\frac{\lambda_{ph}}{\gamma_{ph}}
\frac{(\Omega_{ph}\tau)^2
\left[1-\left((1+\bar{\lambda})(\Omega_{ph}\tau)\right)^2\right]}
{\left[1+\left((1+\bar{\lambda})(\Omega_{ph}\tau)\right)^2\right]^2}\>,
\end{equation}
where
$\tau=\tau(\Omega_{ph})$ and $\bar{\lambda}=\bar{\lambda}(\Omega_{ph})$.
I would like to  stress that $\tau$ and $\bar{\lambda}$ are connected
with {\em all} scattering processes of the carriers and not just
with the electron-phonon scattering. It follows from Eq.~(12) that the
contribution of the process described in Fig.\ 1(b) will produce a ``notch''
in the conductivity near the $LO_{c}$ phonon if $(1+\bar{\lambda})^{-1}
\tau^{-1}>\Omega_{ph}$, and that the relative depth of the notch,
 $\delta\sigma_{1}(\omega=\Omega_{ph})/\sigma_{1}(\omega=\Omega_{ph})$
 is proportional to $\lambda_{ph}\Omega_{ph}/\gamma_{ph}$.
More generally, the condition for a ``notch'' implied by Eq.~(9) is
$\vert Im\Pi(\omega)\vert > Re\Pi(\omega)$ for $\omega$
near the phonon position. At
temperatures $T\geq \Omega_{B}/3$, where $\Omega_{B}$ is
the characteristic energy of a boson which produces a dominant contribution
to the electron self-energy, $\bar{\lambda}$ in Eq.~(12) could be ignored
and one  finds for example that the maximum size of
$\delta\sigma_{1}(\omega=\Omega_{ph})/\sigma_{1}(\omega=\Omega_{ph})$
 is $0.3\lambda_{ph}\Omega_{ph}/
\gamma_{ph}$ when the scattering rate of
the carriers is about $2\Omega_{ph}$.
In the case of the $LO_{c}$ phonons in $Pb_{2}Sr_{2}DyCu_{3}O_{8}$, where
the minima in the room temperature reflectance have been observed between
$400\>cm^{-1}$ and $600\>cm^{-1}$ \cite{Reedyk1},
this would correspond to $1/\tau$
of about $1000\>cm^{-1}$-a value similar to what has been found for the
scattering rate of ``mid-infrared carriers'' in $YBa_{2}Cu_{3}O_{7}$
in the same frequency range \cite{Timusk2}.
\begin{figure}
\centerline{\psfig{file=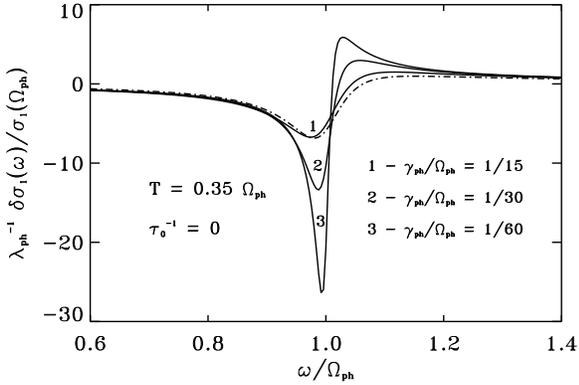,width=0.95\linewidth}}
\caption{$\delta\sigma_{1}(\omega)$  calculated without
impurity scattering for a $LO_{c}$ phonon at frequency $\Omega_{ph}$
and different ratios of $\gamma_{ph} / \Omega_{ph}$.
The temperature $T=0.35\Omega_{ph}$ corresponds to room temperature
if phonon frequency, $\Omega_{ph}$, is equal to 600 cm$^{-1}$.
The parameters for the electron-boson spectral weight are
$\lambda=2$, $\gamma_{B}= 0.25\Omega_{B}$, $\Omega_{B}= 0.66 \Omega_{ph}$
 for
curves 1-3 and $\Omega_{B}= \Omega_{ph}$ for dash-dotted
curve  (see the text).
The coupling constant $\lambda$ is defined as twice the integral
over the electron-boson spectral weight divided by
frequency [8].}
\end{figure}

In order to obtain more quantitative results for $\delta\sigma_{1}$ one
needs both the real and the imaginary part of $\Pi(\omega)$.
The normal state value of $\Pi$ in Eq.~(9) is evaluated in the usual way
\cite{Lee,Bickers,Kostur1}, by replacing it with a simple bubble
$\Pi_{1}$ in Fig.\ 1(b), where the electron lines are dressed by the
self-energy arising from a one-boson exchange and/or from the impurity
scattering. The ladder diagrams involving electron-boson and
electron-impurity scattering processes which contribute to $\Pi_{2}$,
Fig.\ 1(b), can be effectively included in $\Pi_{1}$ by replacing
the electron-boson spectral function and electron-impurity scattering rate
which appear in $\Pi_{1}$ by corresponding transport quantities
\cite{Bickers,Virosztek,Zeyher,Kostur2}.
In Fig.~2 $\delta\sigma_{1}(\omega)$, Eq.~(9),
 is shown for several values of
$\Omega_{ph}/\gamma_{ph}$. The vertical scale is chosen such that it is
dimensionless and such that the results do not depend on the size of the
electron-phonon coupling parameter $\lambda_{ph}$.
The electron-boson
spectral function used in the calculation of ${\Pi}_{1}$
had a form of a wide Lorentzian (half-width $\gamma_{B}$) centered at
$\Omega_{B}$.  The Lorentzian was cut off at $\Omega_{B}\pm 3\gamma_{B}$
\cite{Bickers}. Clearly, the depth of a ``notch'' increases with increasing
sharpness of the phonon, in agreement with our approximate analysis.
However, the coupling parameter $\lambda$ for the electron-boson
spectral weight used to obtain the results in Fig.\ 2 is equal to $2$,
which is much larger than the the values deduced from the dc transport
measurements (i.\ e.\ $\lambda\approx 0.3$-$0.5$).
Similar large values of $\lambda$ were needed in the calculation of the
self-energy effects on Raman-active phonons \cite{Zeyher}.
Smaller values of $\lambda$ will not give $\delta\sigma_{1}$
which is observed in the experiments \cite{Reedyk1} (unless the
impurity scattering rate, $\tau_{0}$
 used in the calculation of $\Pi_{1}$ is assumed to
have an unreasonably large value $\geq 1000\>cm^{-1}$).
This is illustrated in Fig.~3.
\begin{figure}
\centerline{\psfig{file=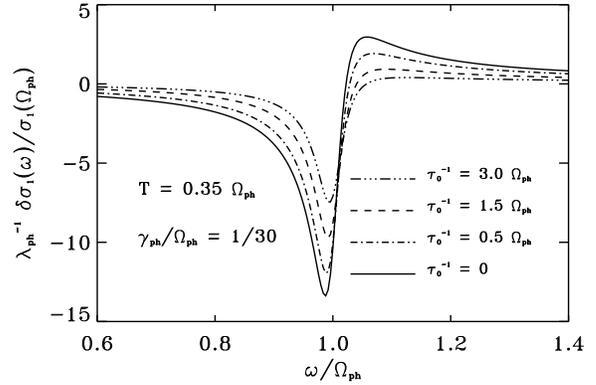,width=0.95\linewidth}}
\caption{$\delta\sigma_{1}(\omega)$ for various
impurity concentration rates $\tau_{0}$. The parameters of the boson
spectrum are
$\lambda=2$, $\gamma_{B}= 0.25\Omega_{B}$, $\Omega_{B}= 0.66 \Omega_{ph}$.
Note that curve 4 is the same as curve 2 in Fig.~2.
The effect of impurity is to suppress the ``notch''-feature in $a$-$b$-plane
conductivity.
}
\end{figure}
One can conclude  that effect of
impurities is to suppress
the ``notch''. The results for
$\delta\sigma_{1}$ shown in Figs.~2 and 3 do not depend on a precise
form of the electron-boson spectral weight which is used to calculate the
dressed electron lines in ${\Pi}_{1}$, as long as the parameter $\lambda$
has a value $\geq 2$ (see dash-dotted line in Fig.~2).
The reason that a large $\lambda$
($\geq 2$) is needed to obtain agreement with experiments \cite{Reedyk1} is
that when $\Pi$ is approximated by $\Pi_{1}$, only these large values of
$\lambda$ will produce large enough scattering rates of the carriers
so that the basic condition for the ``notch''
$\vert Im\Pi(\omega)\vert > Re\Pi(\omega)$ near an $LO_{c}$
phonon is fulfilled. It should
be noted that the fits of $\sigma_{1}(\omega)$
in the mid-infrared range also
require large scattering rates in high-$T_{c}$
materials \cite{Tanner}. A large increase in the scattering rate as one goes
from
far- to mid-infrared range has been interpreted \cite{Tanner} to signal the
existence
of two types of carriers--``Drude'' and ``mid-infrared''.
In Fig.~4 the temperature dependence of $\delta \sigma_{1}$ is shown.
\begin{figure}
\centerline{\psfig{file=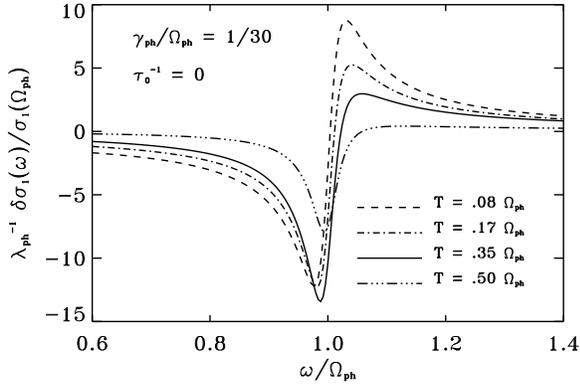,width=0.95\linewidth}}
\caption{Temperature dependence of
$\delta\sigma_{1}(\omega)$ at the set of parameters for
the boson spectrum as in Fig.~3. The strong temperature dependence is
due to approximation of $\Pi$ by simple bubble $\Pi_1$ shown in Fig.~1(b).}
\end{figure}
The ``notch'' feature becomes more pronounced with decrease
in temperature down to temperatures of the order of
$\Omega_B/3 \div 4$;
it then begins to disappear
with further decrease of temperature. For boson frequencies
$\Omega_{B} \sim$ 40 $meV$ this range of temperatures
is close to the superconducting transition temperature $T_c \approx 100$ K.
The peak on the right-hand side of the
``notch'' is related to the real part of the phonon propagator and it is
growing with decrease in temperature
because $Re \Pi$ approaches the value close to $Im \Pi$,
and $Re D$ contributes more (see Eq.~(9)).

It is expected that the mechanism presented here would lead to the same
effects in the superconducting state. For phonons located far above the
peak in the superconducting electron density of states $N_{s}(\omega)$
the effect should be the same  as what we found for the normal state, since
in that case there is no major change in $\Pi(\omega\approx \Omega_{ph})$ due
to the superconducting transition. However, if the phonon is located near
the peak in $N_{s}(\omega)$, as it seems to be the case for $YBa_{2}Cu_{3}O_{7-
\delta}$ \cite{Reedyk1}, the calculation of
$\Pi$ in the superconducting state would
be required.

In conclusion  a microscopic mechanism
which explains a correlation
between the position of $LO_{c}$ phonons and the positions of the
``notches'' in the
$a$-$b$-plane optical conductivity of high-$T_{c}$ copper-oxides, and the
absence of these antiresonant features when the wave vector of light is
perpendicular to the $c$-axis is given.
The puzzling aspects of the experimental work; namely how does the
direction of ${\bf q}^{0}$ play a role, how can light polarized along
{\it a,b} couple to phonons polarized along {\it c} and why is it the
infrared inactive longitudinal modes that are observed, are explained
naturally.  In present  model the light, which is polarized along {\it a,b},
couples to the charge carriers along {\it a,b}; a particle-hole pair is
created which after a series of scattering events through which all
information regarding the original polarization of the light is lost,
recombine emitting a phonon which
{\it can} be polarized along {\it c}.
Whether $c$-axis  phonon can be emitted
depends  on the electron-phonon
coupling matrix element and it is not related
to light scattering.
The  non-zero contribution to
Fano antiresonance-type
 process
 is due to the layered structure
which implies the correlation between momenta of electrons
in different planes (see Eq.~(5-6)).
The modes polarized in plane cannot give the antiresonace behavior
because of zero-result for the average of the product of
electron-phonon
matrix element and  velocity.
 With ${\bf q}^{0}$
 perpendicular to {\it c}
 the transverse {\it c}-axis
 modes
could in principle couple
with non-zero $\lambda_{ll'}^{TO_c} ({\bf k -k'})$
and ${\bf e}_{TO_c}({\bf q}^{0}) \nabla \approx
 \frac{\partial}{\partial z_{l}}$,
but after summation over $l$ the contribution
disappears.
The key physical points of the present
work are that the relative size of the
antiresonant features is determined by
$\lambda_{ph}\Omega_{ph}/\gamma_{ph}$ and
that a large ($\geq 1000\>cm^{-1}$)
scattering rate of the carriers in the region
of $LO_{c}$ phonons is required.
These points do not depend on our present inability to calculate the
correlation function $\Pi$  for copper-oxides from the first principles,
as indicated by too large a value of the electron-boson coupling parameter
$\lambda$ which is required when $\Pi$ is approximated
by a simple bubble. In
principle one could use in Eq.~(9)
$\Pi(\omega)$ determined from the
experimentally measured $Re\sigma$ and $Im\sigma$
in perpendicular geometry and
fit the values of
$\lambda_{ph}$, $\Omega_{ph}$ and $\gamma_{ph}$ in Eq.~(9)
for the phonon propagator to the experimental results for $\sigma_{1}$ in
parallel geometry \cite{Reedyk1}.

{\it Acknowledgement}.
The author  would like to thank B.~Mitrovi\'c for many helpful discussions.
He is  also  grateful to
D. N. Basov, J. P. Carbotte, D.~H. Drew,  R.~A. Ferrell, W.~E. Pickett,
  M. Reedyk,  T. Timusk and A.-M. Tremblay for discussions
and their critical  comments.
This work was supported in part by NASA Grant No. NAG3-1395.

\end{document}